\documentclass[inactive,preprint,tightenlines,superscriptaddress, amsmath,amssymb,aps,prl,notitlepage,11pt]{revtex4-2}

\usepackage{graphicx}

\usepackage{subfiles}
\usepackage{xr-hyper}
\usepackage{array,multirow,graphicx,float,makecell}
\usepackage{nameref}
\usepackage[normal]{threeparttable}
\usepackage[capitalize]{cleveref}
\usepackage{xcolor}
\creflabelformat{equation}{#2[#1]#3}

\usepackage{tablefootnote}

\newcommand{\R}{\mathbb{R}}		
\makeatletter
\newcommand\setcurrentname[1]{\def\@currentlabelname{#1}}
\makeatother

\DeclareMathOperator*{\argmin}{arg\,min}
\DeclareMathOperator*{\argmax}{arg\,max}
\newcommand{\beginsupplement}{
    \setcounter{section}{0}
    \renewcommand{\thesection}{S\arabic{section}}
    \setcounter{equation}{0}
    \renewcommand{\theequation}{S\arabic{equation}}
    \setcounter{table}{0}
    \renewcommand{\thetable}{S\arabic{table}}
    \setcounter{figure}{0}
    \renewcommand{\thefigure}{S\arabic{figure}}
}

\begin{document}
\title{Cell reprogramming design by transfer learning of functional transcriptional networks}

\author{Thomas P. Wytock}

\affiliation{Department of Physics and Astronomy, Northwestern University, Evanston, Illinois 60208, USA.}
\affiliation{Center for Network Dynamics, Northwestern University, Evanston, Illinois 60208, USA.}

\author{Adilson E. Motter} 

\affiliation{Department of Physics and Astronomy, Northwestern University, Evanston, Illinois 60208, USA.}
\affiliation{Center for Network Dynamics, Northwestern University, Evanston, Illinois 60208, USA.}
\affiliation{Department of Engineering Sciences and Applied Mathematics, Northwestern University, Evanston, Illinois 60208, USA.}
\affiliation{Northwestern Institute on Complex Systems, Northwestern University, Evanston, Illinois 60208, USA.}
\affiliation{National Institute for Theory and Mathematics in Biology, Evanston, Illinois 60208, USA.}
\vspace{24pt}
\begin{abstract}
Recent developments in synthetic biology, next-generation sequencing, and machine learning provide an unprecedented opportunity to rationally design new disease treatments based on measured responses to gene perturbations and drugs to reprogram cell behavior. 
The main challenges to seizing this opportunity are the incomplete knowledge of the cellular network and the combinatorial explosion of possible interventions, both of which are insurmountable by experiments. 
To address these challenges, we develop a transfer learning approach to control cell behavior that is pre-trained on transcriptomic data associated with human cell fates to generate a model of the functional network dynamics that can be transferred to specific reprogramming goals.
The approach additively combines transcriptional responses to gene perturbations (single-gene knockdowns and overexpressions) to minimize the transcriptional difference between a given pair of initial and target states.
We demonstrate the flexibility of our approach by applying it to a microarray dataset comprising over 9,000 microarrays across 54 cell types and 227 unique perturbations, and an RNASeq dataset consisting of over 10,000 sequencing runs across 36 cell types and 138 perturbations. 
Our approach reproduces known reprogramming protocols with an average AUROC of 0.91 while innovating over existing methods by pre-training an adaptable model that can be tailored to specific reprogramming transitions.
We show that the number of gene perturbations required to steer from one fate to another increases as the developmental relatedness decreases.  We also show that fewer genes are needed to 
progress along developmental paths than to regress. 
Together, these findings establish a proof-of-concept for our approach to computationally design control strategies and demonstrate their ability to provide insights into the dynamics of gene regulatory networks.
\vspace{12pt}\\
\textbf{Significance summary:} The lack of genome-wide mathematical models for the gene regulatory network complicates the application of control theory to manipulate cell behavior in humans. We address this challenge by developing a transfer learning approach that leverages genome-wide transcriptomic profiles to characterize cell type attractors and perturbation responses. 
These responses are  used to predict a combinatorial perturbation that minimizes the transcriptional difference between an initial and target cell type, bringing the regulatory network to the target cell type basin of attraction. We anticipate that this approach will enable the rapid identification of potential targets for treatment of complex diseases, while also providing insight into how the dynamics of gene regulatory networks affect phenotype.
\vspace{12pt}\\
 Wytock, T. P., \& Motter, A. E. (2024). Cell reprogramming design by transfer learning of functional transcriptional networks. \emph{Proc. Natl. Acad. Sci. USA}. \textbf{121} (11), e2312942121.\\
 \url{https://doi.org/10.1073/pnas.2312942121}
\end{abstract} 

\keywords{biological networks $|$ data-driven control $|$ nonlinear dynamics $|$ cell reprogramming }
\maketitle
\newpage

\section*{Introduction}
The major bottleneck in designing protocols to control cell behavior no 
longer lies with the availability of experimental tools to manipulate 
cellular dynamics, microenvironment, or genetics, but in the ability to 
triage the combinatorial explosion of possible interventions to 
rationally direct experimental efforts.
Advances in synthetic biology are steadily increasing the breadth and 
precision of possible intervention tools, whether they be 
nanoparticles~\cite{morton2014nanoparticle,dasilva2017potential} and 
minicells~\cite{macdiarmid2009sequential} for 
targeted drug delivery, CRISPR and its variants for targeted 
perturbation of the genetic code~\cite{Cong2013} and cellular dynamics~\cite{qi2013repurposing,liu2018crispr}, or 
immunotherapy-based approaches for cancer treatment~\cite{Ludwig2014,McDermott2015,june2018cart}. 

The large corpus of potential interventions and combinations thereof 
make brute-force trial and error approaches too
expensive and time-consuming to be feasible. 
Unlike engineered systems in which control theory provides an 
equation-based framework to design interventions~\cite{Cornelius2013,Wells2015}, 
biological systems are only beginning to attain genome-scale mathematical descriptions~\cite{szigeti2018blueprint}, 
while technical limitations constrain the number of actuable degrees of freedom (genes) to be much smaller than number of components to be controlled.
These features present a challenge given that underactuated control is onerous even in physical systems that admit a closed-form mathematical description~\cite{yu2013survey}.
As a result, the biological control problem is often relaxed to steering between natively stable states~\cite{Muller2011,Yang2016,Zanudo2017,Newby2022},
rather than stabilizing natively unstable ones.
Since transcriptomic measurements are the most frequently employed technique
for querying the cell state, the formulation of data-driven control presented here 
requires only publicly available data
and is robust against the high dimensionality, multi-cell averaging, and low temporal resolution
 typical of these data.
Our goal is to design a general data-driven control approach tailored to these aspects of the data. 
Our approach contrasts with
related control-theoretic formulations in the literature, which
cannot be easily deployed as they usually  require targeted experiments to design the controller~\cite{Baggio2021},
temporally rich data~\cite{Proctor2018}, 
individually resolved system 
trajectories~\cite{Canaday2021,Kim2021}, and/or the availability of microscopic agent-based models~\cite{Patsatzis2023}.
It also contrasts with existing  heuristic approaches to manipulate cell behavior, which can be 
categorized as network-based or annotation-based. 

The network-based approaches  require
an explicit  reconstruction of network interactions~\cite{campbell2016cell,Yang2016,Zanudo2017,Marazzi2022,Newby2022}
and may additionally rely upon a description of the network dynamics~\cite{Steinway2014,zanudo2015cell}.
On the other hand, the annotation-based approaches
focus on whether specific transcription factors are highly expressed
in the target state~\cite{Takahashi2007,dalessio2015systematic,rackham2016mogrify},  
without further considering network interactions.
While each type of approach successfully addresses the problems for which they were conceived,
they possess specific attributes that preclude their direct application to the present problem. 
Network-based approaches may be structural or dynamical.
The structural approaches assume comprehensive knowledge
of the network structure and are valid under a restricted set of dynamical relationships, whereas
the dynamical
approaches require laborious experimental validation, so they
offer high reliability within a limited scope. 
 In contrast, the annotation-based approaches
downplay the role of gene-gene interactions and make qualitative predictions.

The control approach introduced here employs transfer learning on transcriptomic data to retain the strengths of both the network-based
and the annotation-based approaches while addressing their limitations.
Transfer learning entails pre-training on a broad-based dataset followed by the incorporation of application-specific data~\cite{Theodoris2023}. 
In our case, we use broad-based gene expression and bulk RNA-seq datasets 
consisting of observations across a range of unperturbed cell types to pre-train
a machine learning model that maps transcriptional states to cell type.
Pre-training consists of calculating the gene-gene correlation matrix, decomposing the matrix
into eigengenes---combinations of genes that vary approximately independently of one another~\cite{Alter2000,Wytock2019}---and 
selecting the eigengenes that best distinguish cell types.
The eigengene selection is implemented by iteratively selecting the eigengene that minimizes the cross-validation error of a 
distance weighted $k$-nearest neighbors (KNN) model mapping gene expression to cell type, until the error stops decreasing.
The KNN model plays the role of an objective function in our control approach, and the selected eigengenes
capture the functional network of regulatory interactions between genes
without the need to explicitly reconstruct the underlying network of biochemical interactions.
The functional network of regulatory interactions stabilizes cell types, which can be identified as \textit{attractors} of the regulatory dynamics because cells of a given type exhibit stable phenotypes with distinct expression profiles~\cite{Huang2005}. 
Naturally, the empirical evidence for the existence of attractors 
does not require the specification of a dynamical model, which is consistent with our data-driven approach~\cite{Waddington1956,Kauffman1969,Wells2015}.
Because the KNN model maps out regions of the transcriptional state space associated with each cell type, it can be interpreted as estimating the regions of this space in the neighborhood of the corresponding cell type attractors.
These regions can extend beyond the attractors themselves due to stochasticity while nevertheless remaining in each case within the \emph{attraction basin} of the cell type, which is the set of transcriptional states that would deterministically converge to the attractor.

\begin{figure}[tb]
  \centering
  \includegraphics[width=0.48\textwidth ]{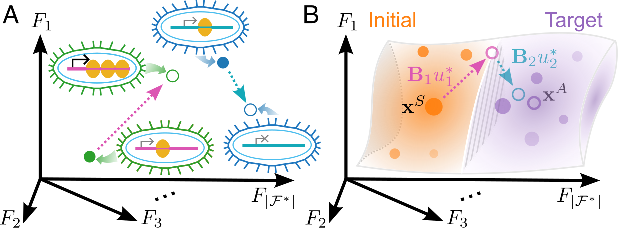}
 \caption{Schematic overview of the data-driven control approach. 
(\emph{A})~Construction of the library of transcriptional responses to gene perturbations in the \emph{latent space}, which is defined as the subspace of selected eigengenes $\mathcal{F}^*$. The pink and teal arrows indicate the experimentally measured shift in transcription from a mock-treated state to a perturbed state (filled and empty circles, respectively) in different cell types (green and blue colors).
(\emph{B}) Perturbation optimization algorithm, where the goal is to drive the initial state $\mathbf{x}^S$ (orange filled circle,  ``S'' for starting) to the target state $\mathbf{x}^A$ (open purple circle,  ``A'' for attractor), which is the average of the individual states of the target cell type  (filled purple circles). 
This is achieved by linearly combining the transcriptional responses to  steer the system to a state (open teal circle) that minimizes the distance to the target. 
Within the algorithm, perturbation responses are added incrementally until the state is predicted to cross the cell type boundary (marked by the patterned surface)
as determined by the KNN model. 
The order in which the incremental perturbations are selected within the algorithm does not imply a temporal ordering in the implementation of the perturbations.
 }
  \label{cartoon:fig}
\end{figure}

Equipped with the pre-trained KNN model, we incorporate transcriptomic data associated with gene perturbations,
which constitutes our application-specific data.
Each gene \emph{perturbation} is either a knockdown or an overexpression (typically of a single gene) and the 
transcriptomic data includes associated mock-treated experiments serving as negative controls. 
\Cref{cartoon:fig}A illustrates the mock-treated (filled symbols) and perturbed states (open symbols) for an overexpression (green) and a knockdown (blue). 
Arrows indicate the \emph{transcriptional response} to the perturbation, defined as the mean difference in expression between the perturbed and mock-treated experiments.
To identify which perturbations can predictively alter 
the transcriptional state from one cell type to another,
we start from unperturbed states of an initial cell type and add the corresponding transcriptional perturbation responses until reaching the basin of attraction of the target
cell type as inferred by the KNN model (\cref{cartoon:fig}B). 
In these predictions, the selected eigengenes are the same for all reprogramming tasks, 
and the selection of perturbations is made application-specific by scaling the projections of the initial-target state distance in the eigengene basis according to the transcriptional variance of the target cell type.

\squeezetable
\begingroup
\begin{table*}[b]
    \centering
    \begin{threeparttable}[t]
    \caption{ \textbf{Statistics of the GeneExp and RNASeq datasets.}}
    \begin{tabular}{ l r r l r r r r
    }
     \hline
     &  Genes, $|\mathcal{G}_D|$  & Eigengenes, $|\mathcal{F}^*_D|$ & Category & Profiles, $N_D$ & Series, $E_D$ &  Cell types, $|\mathcal{C}_D|$& Perturbations, $|\mathcal{P}_D|$ \\[2pt]
      \hline
     \multirow{3}{*}{GeneExp} & 
     \multirow{3}{*}{17{,}525\hspace{3mm}} & 
     \multirow{3}{*}{\ \ 4\hspace{1cm}} 
              & Unperturbed  &  3{,}103\hspace{4mm} & 136\hspace{4mm} & 91\hspace{7mm} & 0\hspace{1cm} \\
       & & & Perturbed & 5{,}735\hspace{4mm} & 356\hspace{4mm} & 368\hspace{7mm} & 207\hspace{1cm} \\
       & & & Reprogrammed & 296\hspace{4mm} & 24\hspace{4mm} & 13\hspace{7mm} & 10\hspace{1cm} \\ \hline
     \multirow{2}{*}{RNASeq} & 
     \multirow{2}{*}{17{,}361\hspace{3mm}} & 
     \multirow{2}{*}{10\hspace{1cm}} 
               & Unperturbed &  9{,}851\hspace{4mm} & 1\hspace{4mm} & 36\hspace{7mm} & 0\hspace{1cm} \\
	& & & Perturbed & 1{,}348\hspace{4mm} & 24\hspace{4mm} & 20\hspace{7mm} & 138\hspace{1cm} \\ \hline
    \end{tabular}
    \label{tab:summary-stats}
    \end{threeparttable}
\end{table*}
\endgroup

\section*{Results}
\label{sec:Results}
\subsection*{Data description}

We apply our approach to human cells using a gene expression microarray dataset (``GeneExp'') and an RNA-sequencing dataset (``RNASeq''),  which are described in \cref{tab:summary-stats}.
Each dataset has a fixed set of measured genes $\mathcal{G}_D$ and a fixed set of selected eigengenes $\mathcal{F}^*_D$, where $D$ labels the dataset. These datasets are partitioned into unperturbed cell states used for training the KNN model, perturbed states used for defining the transcriptional response matrix, and (in the case of the GeneExp dataset) reprogrammed states used for validating the predictions.
The summary statistics for the partitions include the number of experiments $N_D$, the number of experimental series $E_D$, the set of cell types $\mathcal{C}_D$, and the set of perturbations $\mathcal{P}_D$.

\begin{figure}[h]
\centering
  \includegraphics[width=.8\textwidth ]{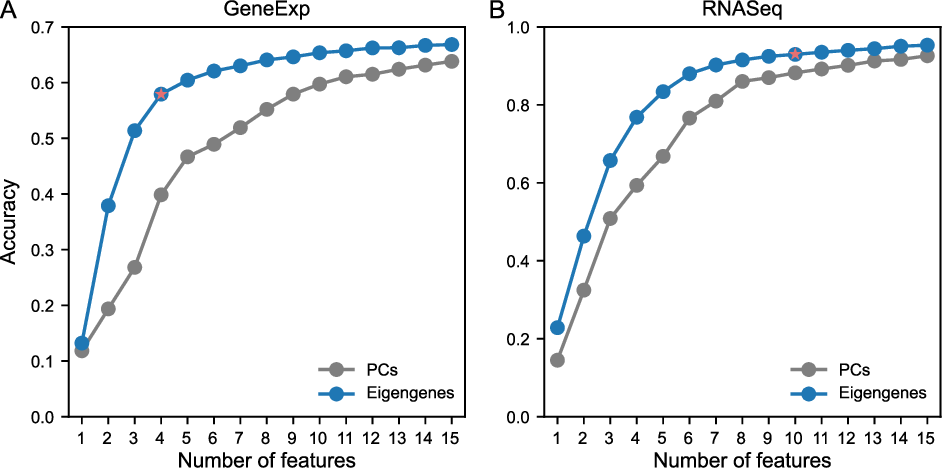}
 \caption{Leave-one-group-out cross-validation accuracy (fraction of instances correctly predicted) of the KNN classification model as a function of the number of features (eigengenes or principal components) for the  (\emph{A}) GeneExp and (\emph{B}) RNASeq datasets. 
Each panel compares the accuracy for a fixed number of eigengenes against the accuracy when using the same number of principal components (choosing those that account for the largest fraction of the variance). Red symbols mark the selected number of features, which is when the accuracy stops significantly improving. This cross-validation was conducted as described in ref.~\citenum{Wytock2020}.
\vspace{-0.3cm}
  }
  \label{fig:logo-cv}
\end{figure}

Each partition serves a distinct role in our approach. 
In both datasets, the \emph{unperturbed} partition consists of cells free from exogenous stimulation, drug treatment, and genetic knockdown or overexpression. These data are used for training our recently developed machine learning method, previously used to distinguish cell type~\cite{Wytock2020}, which selects the optimal set of eigengenes $\mathcal{F}^*_D$.
This method produces a distance-weighted KNN model that maps the latent space to a vector indicating the probability of belonging to each of the $\mathcal{C}_D$ cell types.
Here, \emph{cell type} refers to the phenotypic characterization assigned to the cell sample based on histological and morphological characteristics.
Thus, each transcriptomic measurement in this partition is identified with one cell type, implying that the aspects of the transcriptional state associated with this phenotype are sufficiently long-lived to be considered stable over the timescale of the experiment.
Our KNN method can successfully infer cell type without explicitly reconstructing the regulatory network or the dynamical equations of the system~(\cref{fig:logo-cv}).
The ability of the KNN model to infer the behavior of high-dimensional regulatory networks using a latent space of much lower dimension without losing the relevant biological features may be interpreted as a byproduct of the minimal frustration recently recognized in these networks~\cite{Tripathi2023}.

The \emph{perturbed} partition also applies to both datasets and consists of experimental series, which are sets of experiments associated with the same series-accession number in the Gene Expression Omnibus (GEO) database
and are usually associated with a single study or publication. 
These series have transcriptional measurements of one or more gene knockdowns or overexpressions in addition to associated  mock-treated experiments.
The elements of the set $\mathcal{P}_D$ are metadata identifying the gene and kind of perturbation and are associated with a transcriptional response to that perturbation.
The transcriptional states of genetic perturbations are regarded as steady states that generally persist only as long as the perturbation is induced, implying that the cell type remains unchanged. 
The final transcriptomic measurements of these states are usually taken 24--96 hours after the initiation of the induction.
The transcriptional responses derived from this partition are central to our data-driven control approach described in the next subsection.

The \emph{reprogrammed} partition in the GeneExp dataset consists of experimental series associated with cell reprogramming experiments. 
Since reprogramming is used to refer to several processes in the literature, we clarify that in the remainder of the paper
we exclusively use the term \emph{reprogramming} to refer to the process of transforming differentiated cells into a pluripotent state (i.e., embryonic stem cell-like state capable of redifferentiating into another cell lineage). 
When discussing our results, we use \emph{transdifferentiation} to refer to changing the behavior of a differentiated cell without transitioning through a pluripotent state. 
Compared to perturbation experiments, reprogramming experiments involve more extensive passaging and selection to remove non-responding cells, and the remaining cells do \emph{not} generally return to their original cell type when the perturbation is removed.
The reprogramming experiments in this partition serve as a validation set for our approach.

\subsection*{Data-driven control approach}
\label{subsec:overview}
We now describe how we leverage the partitions of each dataset to arrive at our control approach.
Suppressing the dataset label $D$, we refer to the transcriptional state in the full gene expression space and eigengene space 
using primed $\mathbf{x}'=(x'_i) \in \R^{|\mathcal{G}|}$ and unprimed $\mathbf{x}=(x_i) \in \R^{|\mathcal{F}^*|}$ symbols, respectively.
Revisiting \cref{cartoon:fig}A, each arrow corresponds to a column of the transcriptional response matrix,
$\mathbf{B} = (\mathbf{B}_1, ..., \mathbf{B}_{|\mathcal{P}|}) = (B_{ij}) \in \R^{|\mathcal{F}^*|\times|\mathcal{P}|}$,
represented in the coordinates $\mathcal{F}^*=\{F_1, ..., F_{|\mathcal{F}^*|}\}$.
Our approach to identify perturbations whose transcriptional responses facilitate the transitions between cell types
finds the sum of an initial transcriptional state $\mathbf{x}^S$ and transcriptional responses (i.e., columns of $\mathbf{B}$ scaled by control inputs $\mathbf{u}$) that is as close as possible to the target $\mathbf{x}^A$ (\cref{cartoon:fig}B).
The distance-weighted KNN model $\mathbf{K}(\mathbf{x})$ operates on transcriptional states to infer the probability of cell type membership and assigns a transcriptional state to the most probable cell type---as indicated by the orange and purple background. 
The control inputs $\mathbf{u} = (u_j) \in \R^{|\mathcal{P}|}$ scale the arrows to indicate the extent to which each perturbation is applied.
Specifically, $u_j=0$ means that the $j$th perturbation is inactive, while $u_j=1$ means that this perturbation is active as in the data used to determine $\mathbf{B}_j$.

The control problem in \cref{cartoon:fig}B can be framed as an optimization problem:
\begin{equation}
\mathbf{u}^\dagger = \argmin ||\mathbf{K}(\mathbf{x}^S + \mathbf{B} \mathbf{u}) - \mathbf{K}(\mathbf{x}^A)||_2, \label{eq:opt}
\end{equation}
where  $||\cdot||_2$ is the Euclidean distance between the probability vectors that are output by $\mathbf{K}$. 
The element $u_j$ prescribes the extent to which the $j$th perturbation $\mathbf{B}_{j}$ is active.
We consider three increasingly restrictive scenarios for the control inputs $u_j$: $-\infty<u_j<\infty$, $|u_j|\leq 1$, and $0\leq u_j \leq 1$.
In the first scenario, the control input can alter the initial state to any point on the line $\mathbf{x}^{I} + \mathbf{B}_j u_j$,
while in the second and third scenarios, the range of achievable states is bounded by the magnitude and magnitude \emph{and} direction
of the measured transcriptional response, respectively. 
Moreover, solutions to \cref{eq:opt} that require fewer perturbations are in principle easier to implement experimentally, suggesting a constraint $g=||\mathbf{u}||_0$ ($g$ nonzero elements in $\mathbf{u}$).
These constraints make solving \cref{eq:opt} expensive due to calculating numerical derivatives of $\mathbf{K}$.

To facilitate the identification of experimentally feasible $\mathbf{u}^*$, we approximate
\cref{eq:opt} as 
\begin{equation}
\mathbf{u}^*  = \argmin || \mathbf{x}^S + \mathbf{Bu} - \mathbf{x}^A ||_2. \label{eq:opt_dist}
\end{equation}
\Cref{eq:opt,eq:opt_dist} yield the same solution whenever the nearest neighbor to $\mathbf{x}^S + \mathbf{Bu}^*$ is substantially closer than the $k$th-nearest neighbor (see \nameref{sec:Methods}). 
Note that the approximation in \cref{eq:opt_dist} 
has the advantage of transforming a nonlinear and nonconvex optimization into one that is linear and convex. This is achieved 
by approximating the impact of multiple perturbations as their linear sum
and by approximating differences in KNN-estimated probabilities as differences in transcriptional states.
The former approximation implies that our approach does not temporally order the constituent perturbations within a combination.
The latter approximation is consistent with the observed stability of cell type states,
which guarantees that small perturbations to the observed transcriptional states converge to the same attractor
because otherwise the cell type would be unstable.
Since measurements of a given cell type are in the neighborhood of the same attractor,
the convex hull of the measurements tends to reside within the cell type basin of attraction.
\Cref{eq:opt_dist} can also be expressed as a constrained mixed-integer quadratic program,
 enabling us to take advantage of specialized software. Once $\mathbf{u}^*$ is obtained, $\mathbf{K}(\mathbf{x}^S + \mathbf{Bu}^*)$ is evaluated without approximations to determine whether the target has been reached.

\subsection*{Comparison with existing approaches}
\label{subsec:naive_failure}
We benchmark our data-driven approach against existing approaches to identify candidate reprogramming perturbations
using the $D = {\rm GeneExp}$ dataset.
Approaches that rely on network structure or dynamics are not applicable here due to the lack of a method to generate
predicted transcriptional response from a reconstruction of the gene regulatory network.
The remaining annotation-based approaches select perturbation candidates by compiling lists of genes that are significantly differentially
expressed (DE) between initial and final states~\cite{Takahashi2007,dalessio2015systematic,rackham2016mogrify}. 
These lists are used to identify statistically enriched annotations~\cite{Subramanian2005,Glass2014}, from which differences in pathway regulation and/or transcription factor binding are inferred.
Annotation-based methods have shown promise in  attributing changes to single transcription factors, but we demonstrate here 
that they have limited ability predict the impact of combinations of factors.

\begin{figure}[tb!]
\centering
  \includegraphics[width=.48\textwidth ]{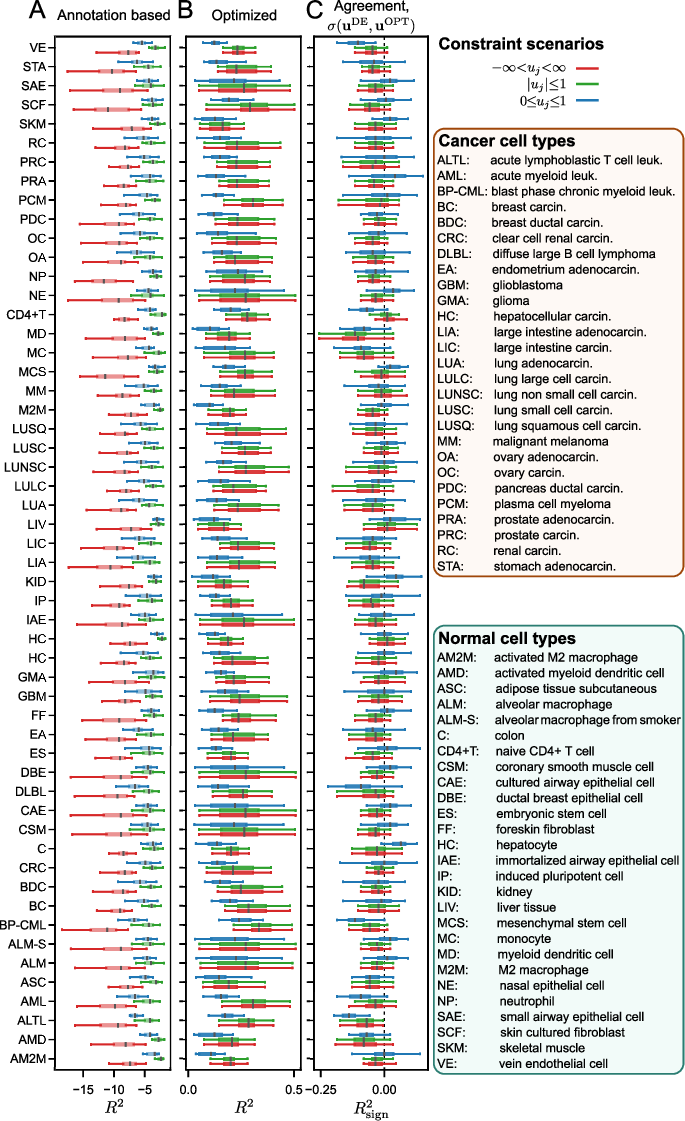}
 \caption{Comparison of annotation-based methods, which do not account for off-target effects, with our control approach, which does. 
 (\emph{A})~Box-and-whisker plots of the coefficient of determination $(R^2)$ of the perturbation predicted using annotation-based methods over all initial cell types for the unconstrained (red), size-constrained (green), and sign-constrained (blue) constraint scenarios applied to each target cell type. 
For each method, the left, center, and right of each box represent the 25th, 50th (median), and 75th percentiles of the distribution, respectively; the whiskers mark
the minimum and maximum, excluding outliers, which are suppressed for clarity.
 (\emph{B})~Results corresponding to those in \emph{A} for our control approach.
 (\emph{C})~Coefficient of determination of the sign of the optimal $u_j$ in each method.
  }
  \label{naive-v-opt:fig}
\end{figure}

\begin{figure}[h]
\centering
  \includegraphics[width=.48\textwidth ]{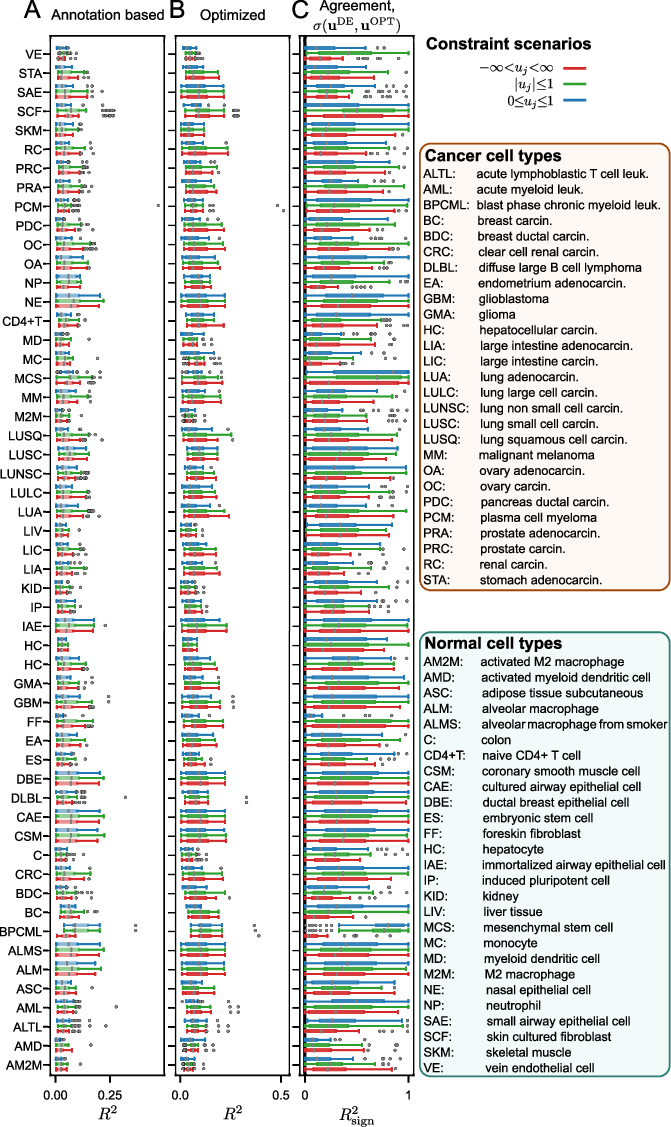}
 \caption{Comparison of annotation-based methods with our control approach when selecting a single gene. 
 (\emph{A}) and (\emph{B}) Boxplot definitions, axes labels, and colors are the same as in \cref{naive-v-opt:fig} in the main text, with the only modification that they represent results for a single-gene perturbation rather than multi-gene perturbations.
 (\emph{C}) Comparison of the distributions of the identities of the single-gene perturbations across all initial states for each method. 
 The extreme values of $0$ and $1$ correspond to the selection of disjoint and coincident sets of perturbations, respectively.
\vspace{-0.3cm}
  }
  \label{naive-v-opt-single:fig}
\end{figure}

We emulate these methods by assigning 
 $u^{\rm DE}_{j} = x'^A_{j} - x'^S_{j}$ for all perturbations that
are measured in the gene expression, i.e., $j \in \mathcal{P}_{D} \cap \mathcal{G}_{D}$.
Using $d(\mathbf{x}^S, \mathbf{x}^A, \mathbf{u}; \mathbf{B})$ to represent the Euclidean distance on the right-hand side of \cref{eq:opt_dist}, we compare $\mathbf{u}^{\rm DE}$ with our method 
$\mathbf{u}^{\rm OPT} =  \argmin_{\mathbf{u}} d(\mathbf{x'}^S, \mathbf{x'}^A, \mathbf{u}; \mathbf{B'}) $
under the three constraint scenarios described above.
This is done using the coefficient of determination
\begin{equation}
R^2(\mathbf{u}; \mathbf{x'}^S\!\!,\mathbf{x'}^A\!\!,\mathbf{B'}) = 1 - d(\mathbf{x'}^S\!\!, \mathbf{x'}^A\!\!, \mathbf{u}; \mathbf{B'}) / d(\mathbf{x'}^S\!\!, \mathbf{x'}^A\!\!, \mathbf{0}; \mathbf{B'}), \label{eq:R2}
\end{equation}
where $R^2 \rightarrow 1$ as $\mathbf{x'}^S + \mathbf{B'u}$ (the sum of the initial state and perturbation responses)  approaches the target $\mathbf{x'}^A$ and $R^2 < 0$ if it is farther away.
For each $\mathbf{x}^S$ in the unperturbed GeneExp partition, we obtain the $\mathbf{u}^{\rm DE}$ and $\mathbf{u}^{\rm OPT}$ using the mean expression of each cell type (different from that of $\mathbf{x}^S$) as $\mathbf{x}^A$.
We then calculate \cref{eq:R2}  for each $\mathbf{u}^{\rm DE}$ and $\mathbf{u}^{\rm OPT}$  and take the median over all states in each initial cell type.
\Cref{naive-v-opt:fig}A and \Cref{naive-v-opt:fig}B, respectively, present these results as box-and-whisker plots across all target cell types for $\mathbf{u}^{\rm DE}$ and $\mathbf{u}^{\rm OPT}$. 
Surprised by the poor performance of the annotation-based methods given their ubiquity in the literature, we 
performed the same analysis for a single gene in \cref{naive-v-opt-single:fig}, 
which shows much
better agreement between these methods and ours.
We infer that the annotation-based method tends to 
move the state farther away from the target (i.e., $R^2<0$) because they have no way to 
calibrate the contributions of each individual perturbation in the sum, since off-target effects are not quantitatively accounted for.
Our optimization-based approach, on the other hand, explicitly accounts for these effects,
and as a result it reduces the initial transcriptional difference by approximately 10\% (when constraining $0\leq u_j \leq 1$) 
to 25\% (in the remaining cases). 
This recovery is 10--25 times larger than the fraction of perturbed genes 
($|\mathcal{P}_D|/|\mathcal{G}_D| \approx 0.01$), which would be the expected recovery for perturbations
that are randomly generated.

 We next investigate whether $\mathbf{u}^{\rm DE}$ agrees qualitatively with $\mathbf{u}^{\rm OPT}$, despite the poor performance of the former. 
This is quantified using the sign alignment metric
\begin{equation}
\sigma(\mathbf{u}^{\rm DE},\mathbf{u}^{\rm OPT}) = \frac{1}{|\mathcal{P}_D \cap \mathcal{G}_D|} \sum_{j=1}^{|\mathcal{P}_D \cap \mathcal{G}_D|} \mathrm{sgn}\left(u^{\rm DE}_j\right) \mathrm{sgn} \left(u^{\rm OPT}_j\right),  \label{eq:sign-align}
\end{equation}
where $\mathrm{sgn}\left(x\right)$ is $1$ if $x>0$, $-1$ if $x<0$, and 0 if $x=0$. 
\Cref{eq:sign-align} is directly applied to  $\mathbf{u}^{\rm DE}$ and $\mathbf{u}^{\rm OPT}$ in the 
first two constraint scenarios  (red and green in \cref{naive-v-opt:fig}),
 but it is applied to $2\mathbf{u}^{\rm DE}-1$ and $2\mathbf{u}^{\rm OPT}-1$ in the $0\leq u_j \leq1$ scenario  (blue in \cref{naive-v-opt:fig}).
 This transformation enables all three scenarios to have equal ranges.
\Cref{naive-v-opt:fig}C shows that the annotation-based perturbations fail to agree with the optimization-based ones qualitatively in terms of the perturbation direction (first two scenarios)
or identity (third scenario). 
Together, these results show that our approach can identify candidate perturbation combinations when annotation-based approaches cannot.

\begin{figure}[t]
  \centering
  \includegraphics[width=85mm]{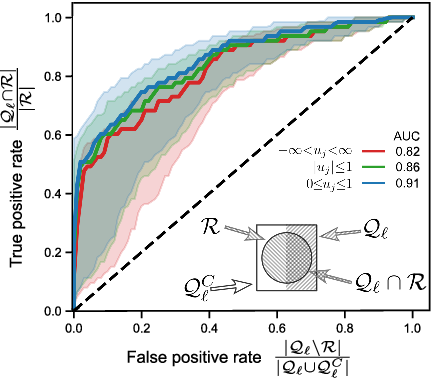}
 \caption{Receiver operator characteristic (ROC) curves demonstrating the ability of our approach to reproduce known reprogramming protocols.
 The ROC curves are constructed by comparing single-perturbation strategies identified by our approach ($\mathcal{Q}$, upper diagonal hatching in the rectangle) ranked in order of their distance to the target (\cref{eq:opt_dist}) against 63 experimentally confirmed reprogramming protocols from the literature ($\mathcal{R}$, lower diagonal hatching in the circle). 
 The sizes of $\mathcal{Q}$ and $\mathcal{R}$ and their overlap are characterized by the true positive rate and false positive rate as defined in the vertical and horizontal axis labels, respectively.
 The color-coded curves and backgrounds correspond to the median and interquartile range for the constraints indicated in the legend, including the
 median area under the curve (AUC).
  }
  \label{fig:reprog_validation}
\end{figure}

\subsection*{Data-driven reproduction of existing protocols}
We next validate our approach by confirming that it can reproduce existing reprogramming protocols.
A protocol consists of a set of perturbations that have been experimentally observed to drive a differentiated
cell type to a pluripotent cell type.
For this and all subsequent analyses, we project the data onto eigengenes and 
estimate cell type using the KNN models.
 Our validation dataset contains of a set of 63 successful reprogramming protocols $\mathcal{R}$ and  220 other perturbations (Fig.~S3).  
We order the set of all perturbations $\mathcal{Q}$ according to the  minimum distance achieved by the optimal single-gene perturbation
under the three constraint scenarios averaged across initial-target pairs.
The initial states  $\mathbf{x}^S$ are drawn from a fixed differentiated
cell type and the $\mathbf{x}^A$ are drawn from a fixed pluripotent cell type in  given GEO series belonging to
reprogrammed partition of the GeneExp dataset.
Using $\mathcal{Q}_{\ell}$ to denote be the first $\ell$ elements of the set of perturbations, 
the true positive rate is $|\mathcal{R} \cap \mathcal{Q}_{\ell}|/|\mathcal{R}|$, and
the false positive rate is $|\mathcal{Q}_{\ell} \backslash \mathcal{R}|/|\mathcal{Q}_{\ell} \cup \mathcal{Q}_{\ell}^C|$, where $\backslash$ is the set difference operator and $\mathcal{Q}_{\ell}^C$ denotes the set complement.
\Cref{fig:reprog_validation} plots the true positive rate as a function of the false positive rate for $\ell \in \{1,...,283\}$
(i.e., the receiver operator characteristic curves) for each constraint case.
As the constraints become more restrictive, the area under the curve increases.
This trend indicates that the restriction of
perturbation strengths to experimentally realizable values improves the identification of viable reprogramming strategies.
Unlike in \cref{naive-v-opt:fig}B, in which the unconstrained and $|u_j|\leq 1$ scenarios produced nearly identical results, 
the $|u_j |\leq 1$ and $0\leq u_j \leq 1$ scenarios produce broadly similar results in \cref{fig:reprog_validation}.
We attribute the difference between the first two scenarios to our approach's ability to quantitatively discriminate between candidate 
reprogramming perturbations based on the magnitude of their transcriptional response (i.e.  $||\mathbf{B}_j||_2$).
In other words, while many single-gene
perturbations have the potential to move the cell state toward the target (pluripotency in this case), 
they have more limited impacts on gene expression 
than the overexpression of the Yamanka factors (KLF4, POU5F1, MYC, SOX2)~\cite{Takahashi2007}.
The similarity of predictive performance between the $|u_j|\leq 1$ and $0\leq u_j\leq 1$ scenarios 
suggests that the $|u_j|\leq 1$ case can be useful for hypothesis generation
in spite of the empirical observation 
that the impact of knocking down a gene is not an exact additive inverse of overexpressing one~\cite{sopko2006mapping}. 
Specifically, genes implicated in the $|u_j|\leq 1$ case correspond to portions of the gene regulatory network that may be targeted by new perturbation experiments to evaluate their utility for reprogramming.

 We also compare our approach against those of an existing method, Mogrify~\cite{rackham2016mogrify}. 
 Unlike our approach, Mogrify does not take the initial state into account and only provides a ranked list of transcription
 factors, with the assumption that all transcription factors are overexpressed. 
 We calculate all single-gene
 perturbations that facilitate reprogramming between the 214 initial cell types in the GeneExp dataset to the 54 tissues
 considered by Mogrify. 
 Using the set of 71 overlapping transcription factors between
 knockouts in the GeneExp dataset and those in Mogrify, we compare the set of transcription factors predicted by our approach  
 with those in the top 1\% for the same target cell type in Mogrify. 
 In all cases, we find at least one overlapping transcription factor between our predictions 
 and those of Mogrify.
 However, we additionally identify other transcription factors in our predictions that reprogram a wider
 range of initial states, demonstrating that taking the initial state into account can generate more state-specific and
 broadly effective reprogramming strategies.

\begin{figure}[bt]
  \centering
  \includegraphics[width=88mm]{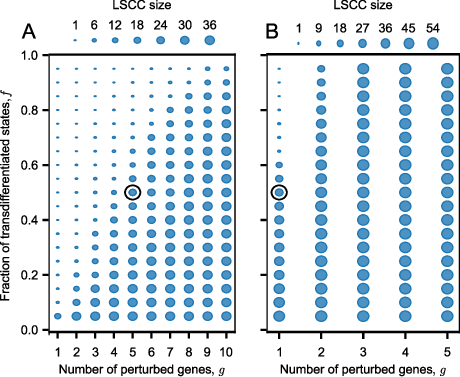}
 \caption{Possible transdifferentiation transitions as a function of the number of  genes perturbed and the fraction of successful transitions. 
 (\emph{A}) Largest strongly connected component sizes of the networks created when including an edge for each initial-target pair in the RNASeq dataset for which at least a fraction $f$ of the initial states (vertical axis) are transdifferentiated using at most $g$ perturbations (horizontal axis).
 (\emph{B}) Corresponding results for the GeneExp dataset.
 The circled cases are considered further in subsequent figures.
  }
  \label{fig:comp_size}
\end{figure}

\subsection*{Analysis of predicted transdifferentiation transitions}
\label{subsec:predict_trans}
We next apply our approach to the GeneExp dataset and the RNASeq dataset
and examine the transdifferentiation strategies it generates in each case.
The states $\mathbf{x}^S$ from initial cell type $s$ are drawn from the unperturbed partition 
in each dataset, while each $\mathbf{x}^A$ is the mean of
all states in the partition belonging to a target cell type $a \neq s$.
Constraining $0\leq u_j \leq1$, we determine the smallest  number of applied perturbations  $g$ that reaches the target basin of attraction for each pair $\{\mathbf{x}^S, \mathbf{x}^A\}$.
We create a transdifferentiation transition network from these results, in which the nodes are cell types and 
edges indicate the  transdifferentiation transitions from $s$ to $a$ that are possible with $g$ or fewer
genes for more than a fraction $f$  of the possible states $\mathbf{x}^S$ in the dataset.
\Cref{fig:comp_size} shows the size of the largest strongly connected component (LSCC) of each transdifferentiation transition network 
as a function of $g$ and $f$ for the (A) RNASeq and (B) GeneExp datasets.
For each value of $f$, there is a number of applied perturbations $g$ for which $g\rightarrow g+1$ results in a rapid increase of the LSCC size, indicating a jump from fragmented subnetworks to a single giant component.  Such a pattern consistent with the hypothesis that few genes are necessary to reprogram cells~\cite{Muller2011}
(a similar trend is observed for small increases in $f$ for fixed $g$).
The LSCC of the GeneExp dataset, for example, 
contains all cell types for only $g=2$ and $f=0.5$, meaning 
that it becomes possible
to steer from one cell type to any other with $g\leq 2$ genes.

\begin{figure*}[t!]
  \centering
  \includegraphics[width=.9\textwidth]{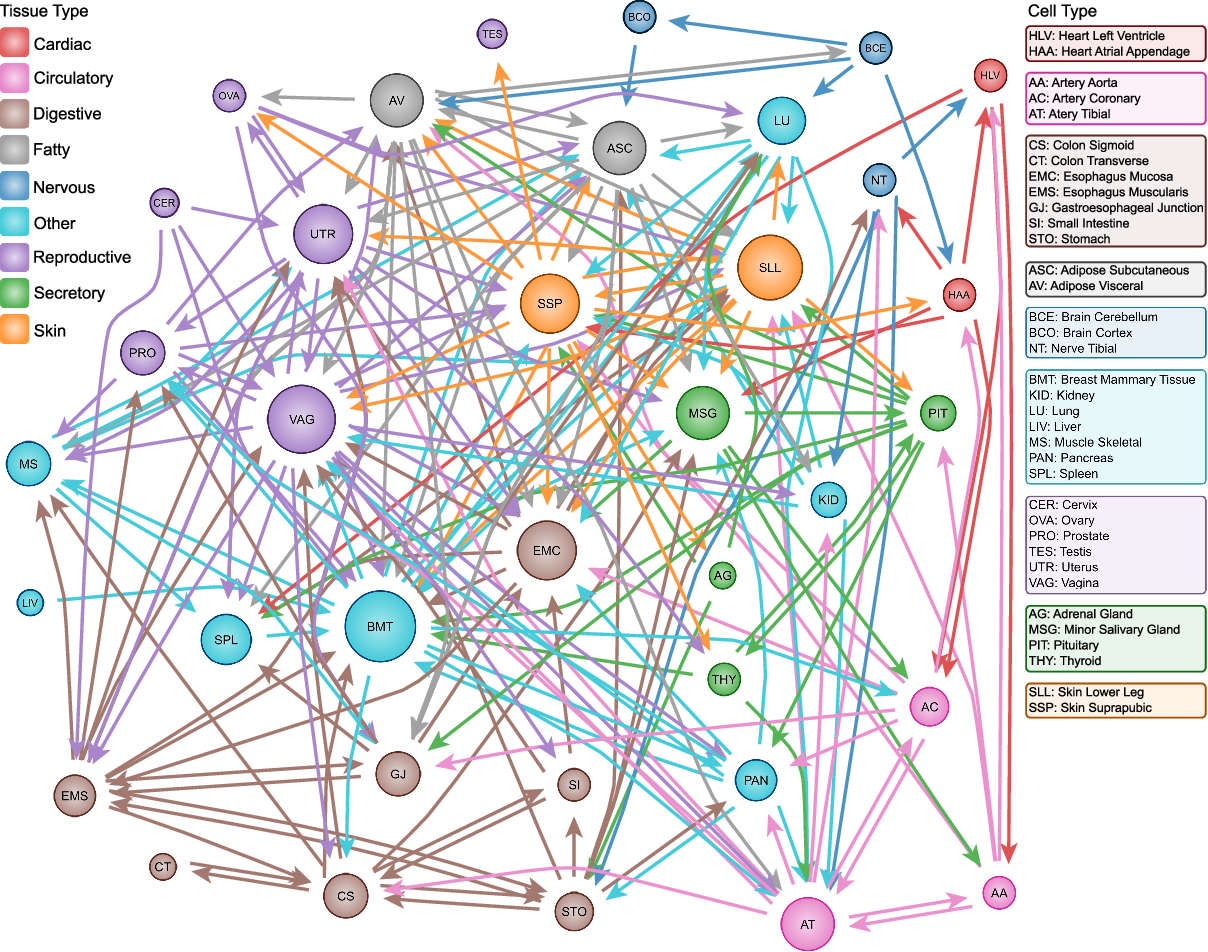}
 \caption{
 Network of transitions (edges) between cell types (nodes) for the parameters indicated by the circle in \cref{fig:comp_size}A.
The nodes and outgoing edges are color coded by tissue type. The node size increases with the total number of edges (i.e., the sum of incoming and outgoing edges).
 }
  \label{fig:RNASeq_reprog}
\end{figure*}

We illustrate the transdifferentiation transition network obtained at $g=5$ and $f=0.5$ 
in \cref{fig:RNASeq_reprog} (the circled instance in  \cref{fig:comp_size}A), which is on the cusp of being fully connected. 
Our approach finds that most transdifferentiation transitions occur between related cell types, in accordance 
with observed developmental patterns.
Cardiac, circulatory, fatty, skin, and to a lesser extent, neurological and digestive tissues illustrate this pattern. 
The main exceptions are to this are reproductive or secretory tissues, which do not seem to preferentially transdifferentiate within their group.

\subsection*{Prominent genes in transdifferentiation transitions} 
In addition to examining the pattern of transitions, we statistically test whether any gene perturbations are 
associated with reaching particular cell types.
For each pair, we test whether the first-selected perturbation 
from a particular initial cell type to a particular target occurs 
more frequently than would be expected by chance. 
The expectation is set by the average of the observed frequencies 
of perturbations from the initial cell type to all other targets. 
Perturbations with frequencies
exceeding this expectation are associated with exiting the initial cell type.
Conversely, perturbations with frequencies exceeding the average 
frequencies observed for transitions into the target cell type from all initial cell types are associated with entering the target cell type (details in \nameref{sec:Methods}).

\begin{figure*}[h]
\centering
\includegraphics[width=\textwidth]{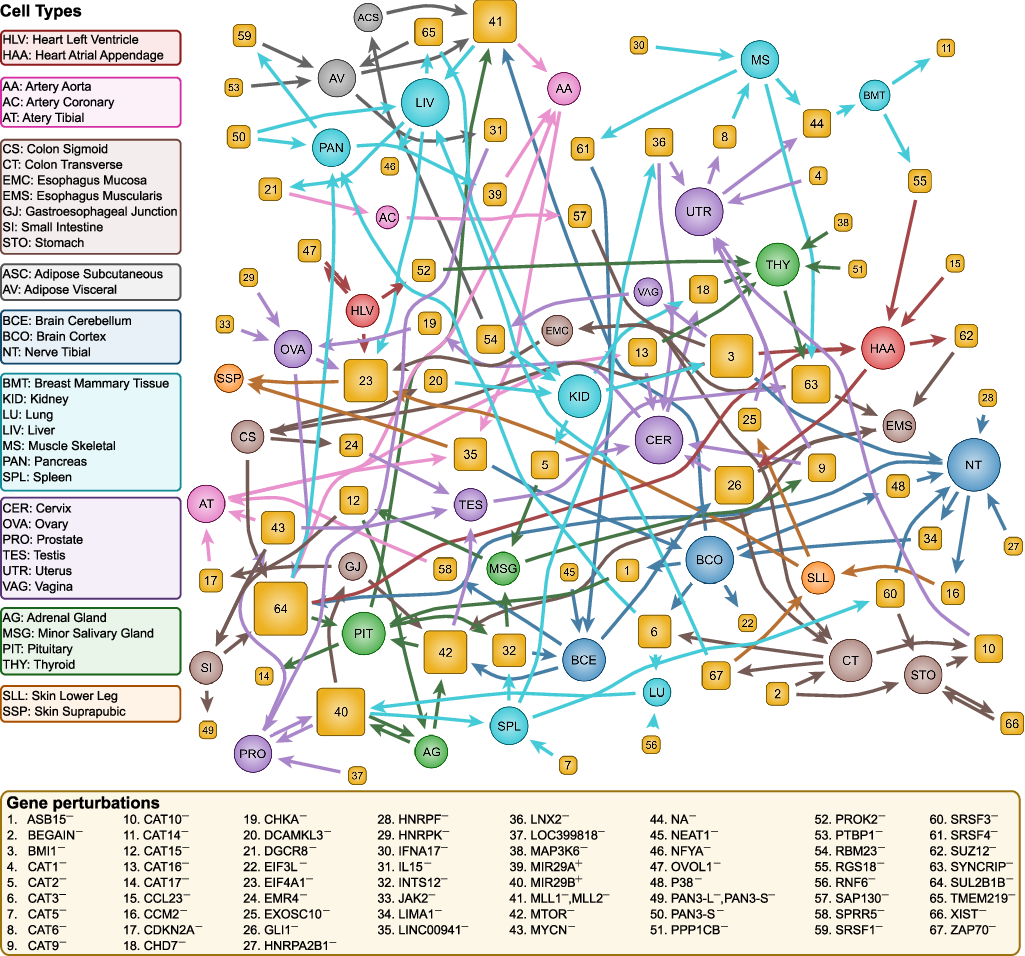}
\caption[Bipartite network of gene perturbations overrepresented in  transdifferentiation transitions for the RNASeq dataset]{
Bipartite network of gene perturbations overrepresented in transdifferentiation transitions for the RNASeq dataset. 
Rectangular nodes, circular nodes, and edges indicate gene perturbations, cell types, and transdifferentiation transitions, respectively.
Circular node colors are as in Fig.~5.
The network consists of genes assessed to be statistically significant with respect to a multinomial null distribution with
a Benjamini-Hochberg multiple hypothesis correction (full details in Methods).}
\label{fig:rnaseq_siggenes}
\end{figure*}

 We present these perturbations and associated cell types in \cref{fig:rnaseq_siggenes}. 
 While few perturbations appear to be associated with specific cell types, we find that 
 digestive cell types share the long non-coding 
 RNA (lncRNA) chromatin associated transcript 10 (CAT10)~\cite{ray2016cat7}, SYNCRIP, SULT2B1, and BEGAIN
 knockdowns between two digestive cell types.
 Furthermore, fatty tissues shared the double knockdown of lysine methyltransferases MLL1 (KMT2A) and MLL2 (KMT2D)
 and arterial tissues shared the lncRNA LINC00941.
 The prevalence of lncRNAs associated with transdifferentiation is consistent with recent experiments establishing the role of these
 factors in determining cell fate~\cite{Sarropoulos2019}.
 Interestingly, the knockdown of the translation initiation factor eIF4A1, which has been suggested to upregulate 
 expression of oncogenes~\cite{modelska2015malignant}, appears to facilitate the departure from the 
 lower-leg skin to the suprapubic skin, which highlights the potential of using gene knockdowns
 to mitigate tissues that have accumulated damage.

\begin{figure*}[t!]
  \centering
  \includegraphics[width=.95\textwidth ]{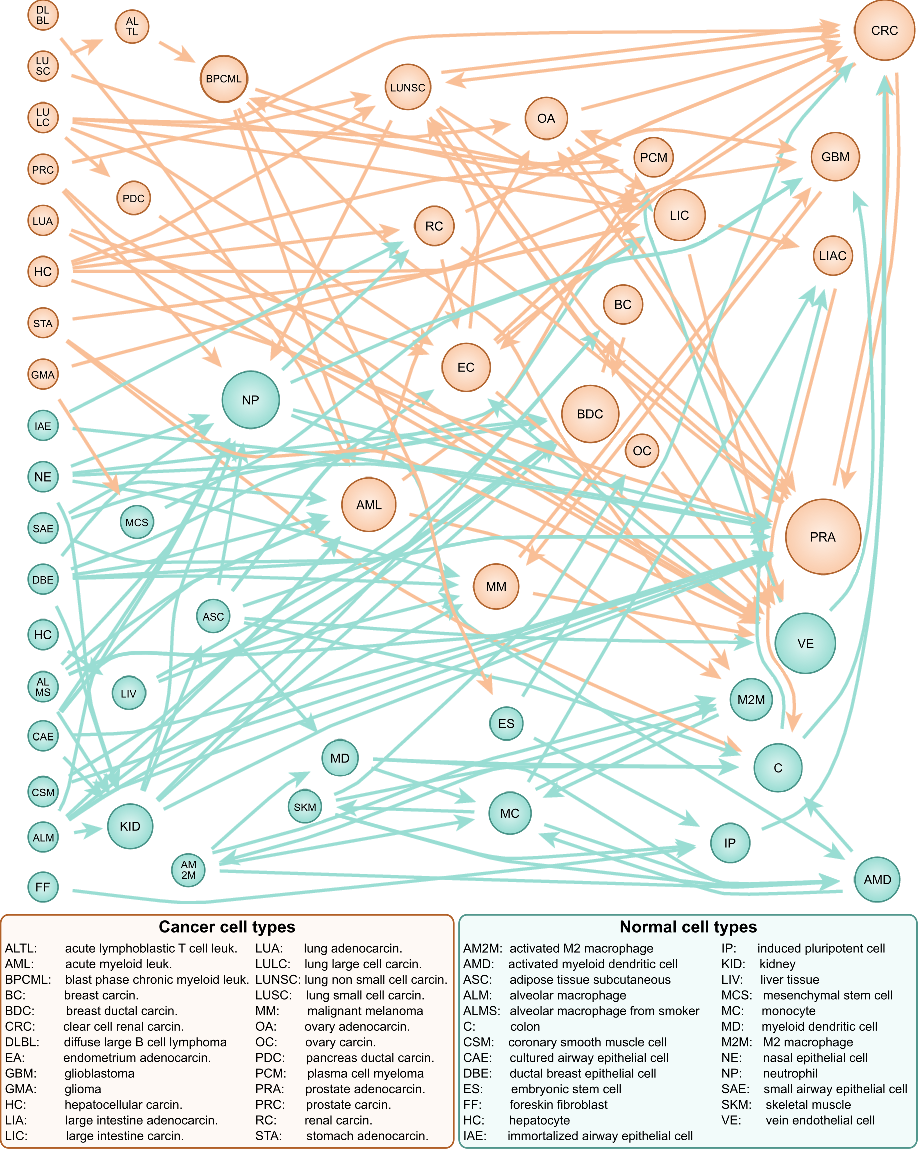}
 \caption{ Network of transitions (edges) between cell types (nodes) for the parameters indicated by the circle in \cref{fig:comp_size}B. 
The number of upstream cell types for a given node increases from left to right, with nodes with zero incoming edges appearing on the left.
 The node size increases with the number of incoming edges and the node color encodes normal (teal) and cancerous (orange) cell types. 
  }
  \label{fig:geneexp_reprog}
  \vspace{-12pt}
\end{figure*}

\Cref{fig:geneexp_reprog} diagrams the pattern of  transdifferentiation transitions
in the GeneExp dataset, which contains a number of normal and cancerous tissues.
Specifically, we observe that cancerous states tend to be reachable from normal states by a single gene, but not vice-versa. 
Of the 25 normal states, only 5 
lie downstream of a cancerous cell type. 
In contrast, 12 of 29 cancerous cells are downstream of normal cells. 
These results are consistent with the observation that cancers tend to arise spontaneously but rarely resolve spontaneously. 
In \cref{fig:geneexp_siggenes}, 
the equivalent to \cref{fig:rnaseq_siggenes} 
for the GeneExp dataset,
we find that the Yamanka factors play a central role in transdifferentiating between cell types.
In addition, the prominence of multiple micro-RNA (MIR31\textsuperscript{+}, MIR34A\textsuperscript{$-$}), mechanotransduction (CDHR2\textsuperscript{$-$}, TWST1\textsuperscript{+}, VEGFC\textsuperscript{$-$}), and metabolic  (IDH1\textsuperscript{$-$}, IDH2\textsuperscript{$-$}, ALDH1A1\textsuperscript{$-$}, ALDH3A1\textsuperscript{$-$}) gene perturbations
is consistent with the observed interplay between cancer progression, mechanosensitivity~\cite{Sullivan2023}, and metabolic reprogramming~\cite{Galbraith2023}.

\begin{figure}[bt]
  \centering
  \includegraphics[width=.48\textwidth]{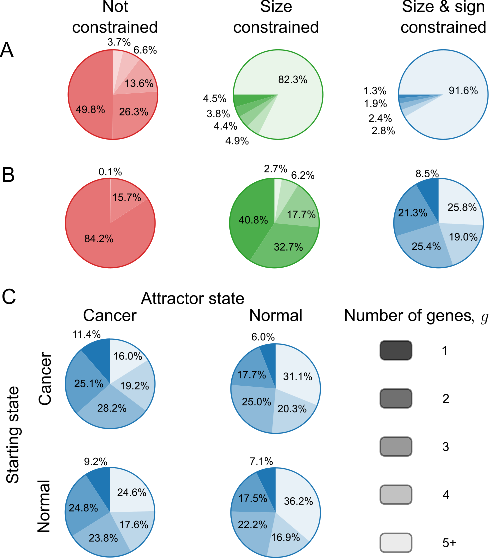}
 \caption{Comparison of transdifferentiation transitions based on the number of genes required. 
 (\emph{A}) Color-coded fraction of  directed cell type pairs in the RNASeq dataset that are able to be transdifferentiated as a function of the number of genes perturbed for all three constraint scenarios. 
 (\emph{B}) Same as \emph{A} but for the GeneExp dataset.
 (\emph{C}) Number of genes required for transitions as a function of the cell type class of the initial state (rows) and target state (columns) for the size-and-sign constrained case in \emph{B}. 
  }
  \label{fig:pies}
\end{figure}

\subsection*{Analysis of the required number of gene perturbations}
Having analyzed the networks of transdifferentiation transitions under the most restrictive case,
we consider the impact of increasing $g$ and relaxing constraints on $\mathbf{u}$. \Cref{fig:pies}~A~and~B shows the fraction of transitions possible as a function of $g$ for the three constraint scenarios.
In the RNASeq (GeneExp) dataset, the fraction of transitions requiring only a single gene increases 3.5-fold (4.8-fold) when relaxing the sign constraint and 38-fold (9.6-fold) when relaxing all constraints.
The more dramatic increase in \cref{fig:pies}A compared to \cref{fig:pies}B
may reflect the greater precision of the RNASeq data, which is also reflected in the larger $|\mathcal{F}^*|$.
\Cref{fig:pies}C shows that the number of genes needed to transdifferentiate from normal to cancerous states
is larger than for the reverse for the constraints $0\leq u_\ell \leq 1$ in GeneExp dataset, quantifying 
the pattern illustrated in \cref{fig:geneexp_reprog}.
These findings demonstrate that our control approach can not only predict candidate genes, but also offer 
a new framework for interpreting the relative stability of cell types. 
Indeed, stability can be characterized by the number genes that need to be 
perturbed to reach or to leave a given cell state, forging a connection with studies of cell type stability in the context of Boolean
reconstructions of regulatory networks~\cite{Joo2018}.

\section*{Discussion}

The results show that our control approach offers 
advantages over both
annotation-based approaches and network-based approaches 
because it improves the quantitative predictive power of the former
and reduces the effort required by the latter to adapt to new systems. 
In particular, the approach has the ability to circumvent the problem of combinatorial
explosion in the number of multi-target interventions.
This is achieved by using the transcriptional difference between states to
computationally triage the combinations of single-target perturbations that are best suited to achieve a particular biological goal, 
whether that be inducing multipotency and pluripotency in differentiated tissues, 
driving transitions between differentiated cell types, or mitigating the progression of cancer.

The approach makes two central assumptions: 
(i) the transcriptional state is the main determinant of cell behavior, 
and (ii) the transcriptional responses to perturbations add approximately linearly to the transcriptional state. 
An important quality of the resulting model is that it can be trained on single-perturbation transcriptional data, which facilitates convergence due to its abundance and coverage of the space of potential perturbations.
Support for (i) follows from our demonstration that it is possible to construct an accurate mapping 
from transcription to cell behavior (Fig.~S1).
Since the approach is based on genome-wide data, it necessarily 
relies on destructive measurements~\cite{Stegle2015}. 
This fact imposes a tradeoff between the depth and time-resolution of the data required for any machine learning method and motivates our focus on long term dynamics.

 Concerning assumption (ii), our approach also serves as a starting point to investigate the role
of nonlinearity at the scale of the whole cell, since 
it evaluates a linear approximation of known responses, and thus
strong deviations from it would be evidence that nonlinear mechanisms are at work. 
Comparison of our approach with deep learning models based on variational autoencoders (VAEs)~\cite{Lotfollahi2019,Lotfollahi2023} reveals that both offer comparable estimates of the mean expression of the final transcriptional state after application of a perturbation (Table~S1). 
This similarity in performance is surprising given that VAEs can in principle learn nonlinear behavior. 
One interpretation of this result is that cells are organized into mostly independent modules whose responses to disparate perturbations are independent and thus combine mostly linearly (as proposed for \textit{E. coli} \cite{Sastry2019}). 
 Indeed, the prevalence of pairwise nonlinear interactions as indicated by statistically significant epistasis is about 7\% in \textit{E. coli}~\cite{Babu2014} and 4\% in \textit{S. cerevisiae}~\cite{costanzo2016global}, with most interactions organized by their genes' functional modules~\cite{segre2005modular}.
If these trends apply to human cells, it could explain the limited amount of nonlinearity seen in previous applications of VAEs~\cite{Lotfollahi2019,Lotfollahi2023}. 
Moreover, nonlinearity is expected to be less pronounced in bulk averages over many cells than in the phenotype of individual cells.

 Bulk transcriptomic data has the advantages that rare transcripts can be detected and that 
histological and morphological observations can be used to supervise the learning of cell type from transcriptional data.
While this limits the ability to detect single-cell heterogeneity~\cite{Stegle2015},
we highlight potential applications of the approach that minimize or account for the impact of cell heterogeneity.
In particular, the approach is suitable for identifying candidate gene perturbations to substitute for potentially oncogenic transgenes when creating na\"{i}ve stem cells~\cite{hou2013pluripotent,bates2017reprogramming}.
It can also be relevant for the management of diseases in which healthy tissues can be treated with gene and/or drug perturbations in \textit{ex vivo} culture 
before autologous re-transplantation~\cite{Ludwig2014}.
The interventions designed by the approach need not be permanent if the target cell state is stable, and they need not be applied to all cells if the modified cells can be selected to out-compete unmodified ones~\cite{McDermott2015}.
Moreover, the approach can be tailored to precision medicine applications by incorporating transcriptional data from individual patients' healthy and diseased tissues to identify treatments
that account for differences between individuals.
These examples illustrate the potential of the approach to computationally screen for effective  regenerative therapies~\cite{jopling2011dedifferentiation}.

Finally, we note that the approach can readily incorporate forthcoming transcriptomic data, 
be applied to modalities other than transcriptional, 
and take advantage of rapidly advancing innovations in machine learning.
The algorithms are designed to incorporate new transcriptomic data without recalculating the latent space, which enables them to capitalize on 
the exponentially increasing abundance of sequencing data~\cite{leinonen2010sequence,Barrett2013}---another 
advantage over knowledge-based control approaches that require a specific dynamical model.
 The versatility of the approach with respect to data modalities 
is important because recent research shows the effectiveness of data on complementary attributes of cell state, such as chromatin accessibility,
in identifying key transcription factors for reprogramming~\cite{Hammelman2022}.
In particular, the approach is amenable to the incorporation of deep transfer learning~\cite{Chen2022}, 
in which deep neural networks could be used to transfer knowledge across data modalities.
Given its many uses and possible extensions, our approach has the potential to become a standard tool to translate bioinformatic data into biomedical applications.
\vspace{-6pt}
\section*{Methods}\setcurrentname{Methods}\label{sec:Methods}
\subsection*{Acquisition of the training data}
The summary statistics for each dataset whose acquisition described below
are provided in \cref{tab:summary-stats}.
The RNASeq dataset consists of (i) unperturbed cell type data
and (ii) gene perturbation data. Part (i) was
obtained from the
GTEx consortium \url{https://www.gtexportal.org/home/} (access date: 09/24/2019),
while part (ii) was curated by searching 
BioProjects from the Sequencing Read Archive (SRA)~\cite{leinonen2010sequence} using the search terms ``crispri'' and
``knockdown'' and retaining the top 40 largest projects (in terms of number of sequencing runs).
Specific details regarding each profile in this dataset, including the accession numbers,  are provided in \cref{tab:RNAseq_pert}.

We constructed the GeneExp dataset using 
human gene expression data from the GEO repository~\cite{Barrett2013}, 
restricting to the most common platform, 
Affymetrix HG-U133+2 microarray (platform accession: GPL570),
to facilitate the comparison of transcriptomic data from different GEO Series of Experiments (GSEs). 
The GeneExp dataset comprises
three parts: (i) unperturbed cell type data to train the KNN classifier,
(ii) gene perturbation data to characterize the corresponding transcriptional response,
and (iii) cell reprogramming data used to validate our method.
We obtained part (i) by searching GEO for the names of the NCI-60 cell lines and obtaining
the relevant data from the GSEs, while supplementing
this with data from the Cancer Cell Line 
Encyclopedia (GSE36139)~\cite{Barretina2012} and the Human Body Index (GSE7307) to gain a representation 
of normal and cancerous cells. 
We curated part (ii) by querying GEO for
``overexpression,'' ``knockdown,'' and ``RNA-interference'' followed by selecting
GSEs that measured gene perturbations.
Finally, part (iii) was found by searching for ``reprogramming,'' yielding 52
different protocols to de-differentiate cells toward a pluripotent state across 18 different
GSEs. 
Specific details relating to each expression profile in this dataset, including the relevant accession numbers,
are provided in \cref{tab:gexp_exps}.

\subsection*{Processing of the expression and sequencing data}
We performed batch correction on the gene expression data using covariates based on experimental series, 
cell line, and experimental treatment to remove systematic variation between series~\cite{Johnson2007}, 
as described in ref.~\citenum{Wytock2020}.
The data and covariates are described in \cref{tab:VAE_comp}
for the RNASeq data and 
\cref{tab:gexp_exps} 
for the gene expression data. 
We used a custom Chip Definition File (CDF) that best maps the probes to the genes~\cite{Dai2005}. 
For the RNASeq data, we used the Ensembl gene identifiers (version: GRCh38.p13) for protein coding genes that overlapped with
those in the GeneExp dataset.

\subsection*{Defining perturbation responses}
To facilitate discussion of the calculation of the cellular response to perturbations, let
$X_{ij}$ be the data matrix of gene expression or transcript measurements, where $i \in \{1,...,|\mathcal{G}|\}$
is an index over genes, and $j \in \{1,...,N\}$ is an index over experiments.
The dataset label $D$ is suppressed to simplify notation.
Let $k$ be an index over GSEs, let $m$ be an index over cell types and culture conditions, let $p$ be
an index over perturbation conditions, and let $\tau$ be the time the sample was collected.
Then, $\mathcal{R}_{(k,m,p,\tau)} \subset \{1,...,N\}$ is the set of columns that 
shares these experimental conditions, and we will refer to the submatrix that
shares these covariates using $\mathbf{X}_{(k,m,p,\tau)}$.
The data will be averaged over these experimental covariates in the following 4 steps:
\begin{enumerate}
    \item[1.] Average the expression over replicates, 
    \begin{equation}
    \bar{\mathbf{X}}_{(k,m,p,\tau)} = \!\!\!\!\!\!\!\! \sum_{j \in \mathcal{R}_{(k,m,p,\tau)}} \!\!\!\!\!\!\!\! \mathbf{X}_{(k,m,p,\tau)} \Big/ |\mathcal{R}_{(k,m,p,\tau)}|. \label{eq:avg_over_rep}
    \end{equation}
    \item[2.] Average over time points,
    \begin{equation}
    \langle \bar{\mathbf{X}}_{(k,m,p)} \rangle = \!\!\!\! \sum_{\tau \in \mathcal{T}_{(k,m,p)}} \!\!\!\!\!\!\!\!\tau \bar{\mathbf{X}}_{(k,m,p,\tau)} \Big/ \!\!\!\!\sum_{\tau \in \mathcal{T}_{(k,m,p)}} \!\!\!\!\!\!\!\! \tau, \label{eq:avg_over_time}
    \end{equation}
    where $\mathcal{T}_{(k,m,p)}$ is the set of time points for the experimental conditions indicated by $(k,m,p)$.
    \item[3.] Restricting to genetic perturbations $p \in \mathcal{P}$ and their controls $0$, calculate differences
    \begin{equation}
    \bar{\mathbf{B}}_{(k,m,p)} = \langle\bar{\mathbf{X}}_{(k,m,p)} \rangle - \langle \bar{\mathbf{X}}_{(k,m,0)} \rangle. \label{eq:pert_diff}
    \end{equation}
    \item[4.] Average over GSEs, cell types, and culture conditions,
    \begin{equation}
    \mathbf{B}_{\ell} = \mathbf{B}_{(p)} = \sum_{q \in \mathcal{P}} \bar{\mathbf{B}}_{(k,m,p)}  \delta_{pq} \Big/ \sum_{q \in \mathcal{P}}  \delta_{pq}, \label{eq:avg_over_conds}
    \end{equation}
    where $\ell \in \{1,...,|\mathcal{P}|\}$ is an index over perturbations and $\delta$ is the Kronecker delta.
\end{enumerate}
\Cref{eq:avg_over_time} weights later time-points more heavily so as to better estimate the long-term response
to the gene perturbation. 
The responses $\mathbf{B}_{\ell}$ are likely to be causal, because they are the outcome of a controlled
experiment, rather than merely correlative. 

\subsection*{Approximating $\mathbf{K}$ with transcriptional distance}
\label{subsec:Sim_Metric}
Our goal is to find the optimal perturbations $\mathbf{u}$ that steer from 
the initial state $\mathbf{x}^S$ belonging to cell type $s$ to the target state $\mathbf{x}^A$ belonging to cell type $a$, as stated in \cref{eq:opt}.
  We recall that $\mathbf{K}$ is the KNN mapping from transcriptional states to cell types, and we have suppressed the dataset labels to simplify notation.
Direct solution of \cref{eq:opt} is challenging because $\mathbf{K}$ is poorly behaved far from
the data used to train it (as discussed below), making methods based on numerical derivatives too slow to employ
due to the computational expense of evaluating $\mathbf{K}$. 
Here, we show that \cref{eq:opt_dist} is an appropriate approximation of \cref{eq:opt} 
under the condition
that $d_k \gg d_1$, where $d_k$ and $d_1$ are the distances to the $k$th-nearest and nearest neighbor in the training data to a test point, respectively.
Let $\mathcal{B}_\varepsilon (\mathbf{x}^A)$
be a ball of radius $\varepsilon>0$ centered at $\mathbf{x}^A$
and define $P(\mathcal{B}_\varepsilon (\mathbf{x}^A))$ to be the probability that $\argmax \mathbf{K}(\breve{\mathbf{x}}^A) = \argmax \mathbf{K}(\mathbf{x}^A)$ over all $\breve{\mathbf{x}}^A \in \mathcal{B}_\varepsilon (\mathbf{x}^A)$. 
Then, 
$\lim_{\varepsilon\rightarrow 0} P(\mathcal{B}_\varepsilon (\mathbf{x}^A)) = 1$ because 
in this neighborhood, the magnitude of the possible discontinuity in 
$\mathbf{K}$ (caused by the change in the $k$th neighbor) is $\varepsilon d^{-1}_k / \left( 1+ \varepsilon \sum_{i=2}^k d^{-1}_i \right)$, which vanishes in this limit. 
As a result, both \cref{eq:opt} and \cref{eq:opt_dist} provide the same answer at infinitesimal distances.
To extend this approximation to finite distances,
we note that (i) the method used to select eigengenes ensures that points within the convex hull 
of the $\mathcal{H}=\{\mathbf{x}^A\, |\, \mathbf{K}(\mathbf{x}^A) = \delta_{ja}\}$ belong to target cell type $a$ with a 
high probability~\cite{Wytock2020} and (ii) the target states we consider are averages of all expression profiles of target cell type $a$, 
$\bar{\mathbf{x}}^A$, which is contained within the convex hull. 
Thus, there is a finite distance $d_\mathcal{H}$ to the nearest boundary of the hull for which 
$P(\mathcal{B}_{d_{\mathcal{H}}} (\bar{\mathbf{x}}^A)) \approx 1$. 

\subsection*{Calculating  transdifferentiation transitions}
The solution of \cref{eq:opt_dist} is underdetermined if the number of control inputs $|\mathcal{B}_D|$ is less than the number of features $|\mathcal{F}_D|$. 
In the underdetermined case, if the control variables are unconstrained, \cref{eq:opt_dist} is solvable by using the Moore-Penrose psuedoinverse of $\mathbf{B}$.
If $|\mathcal{B}_{D}| > |\mathcal{F}_D|$, we impose that the $\ell_2$ norm of the solution $||\mathbf{u}||_2$ is minimized to obtain a unique solution.
The constrained problems reduce to the following program
\begin{equation}
\begin{aligned}
     \argmin_{\mathbf{u}} \quad & d(\mathbf{x}^S, \mathbf{x}^A, \mathbf{u}; \mathbf{B}),  \\
     \operatorname{s.t.}\,\, \quad & l  \leq  u_\ell \leq  1,
\end{aligned}
\label{eq:program}
\end{equation}
where $l=-1$ for the size-constrained case and $l=0$ for the size-and-sign-constrained case,
which is solved using IBM ILOG CPLEX (12.10.0.0).

\subsection*{Limiting the number of genes perturbed}
\label{subsec:Quant_Int}
Since it is experimentally infeasible to target more than a few genes simultaneously, we employ a 
forward selection approach to construct the perturbations. 
Let $\mathcal{V}_{1} = \{ \{j\},\ j \in \{1,...,|\mathcal{B}_{D}|\} \}$ be the set of all single-gene perturbations. 
 Using $u^*_{v_{1}}$ to denote the input that minimizes $d$ for the transcriptional response of column matrix $\mathbf{B}_{v_{1}}$, we compute \cref{eq:opt_dist} by evaluating $d(\mathbf{x}^S,\mathbf{x}^A,u^*_{v_{1}};\mathbf{B}_{v_{1}})$ for each $v_{1} \in \mathcal{V}_{1}$,  and identify the column for which $d$ is minimized, $v^*_{1}$.
We proceed iteratively by constructing the set of all $g$-gene perturbations that include the best $g$--$1$-gene perturbation, denoted $\mathcal{V}_{g} = \{ v^*_{g-1} \cup \{j\},\ j \in \{1,...,|\mathcal{B}_{D}|\}\backslash v^*_{g-1} \}$.
We again evaluate $d(\mathbf{x}^S,\mathbf{x}^A,\mathbf{u}^*_{v_{g}};\mathbf{B}_{v_{g}})$ for all elements $v_{g} \in \mathcal{V}_{g}$ and identify the element $v^*_{g}$ that minimizes $d$. 
We note that $\mathbf{u}^*_{v_{g}}$ is a now a \textit{vector} (as indicated by the bold typeface) of control elements associated with $v_g$, 
which corresponds to the input values that minimize $d$.
Accordingly, $\mathbf{B}_{v_{g}}$ is a \textit{matrix} given by the $g$ columns of the transcriptional response matrix associated with the elements of $v_g$.
We continue this process, incrementing $g$ until the target cell type is reached, i.e.,  $\argmax \mathbf{K}(\mathbf{x}^S + \mathbf{B}_{v^*_{g}} \mathbf{u}^*_{v^*_{g}}) = \argmax \mathbf{K}(\mathbf{x}^A)$. 

\subsection*{Identifying significant genes}
\label{subsec:sig_genes}
We identify significantly overrepresented genes by comparing the frequency of a gene's participation in a specific transdifferentiation transition with
its frequency across all transitions for a given initial cell type $s$ or target cell type $a$.
For the RNASeq dataset, let $\widetilde{N}$ and $|\widetilde{\mathcal{C}}|$ be the numbers of states and cell types in the unperturbed partition,
consider set of optimal control inputs $\mathbf{u}^{(j)}$, where $j \in \{1,...,\widetilde{N}(|\widetilde{\mathcal{C}}|-1)\}$, that are obtained for all pairs of initial states to all target cell types in this partition. 
The functions $I(j)$ and $T(j)$ map the index $j$ to the initial cell type and  target cell type, respectively.
Furthermore, the number of solutions $\mathbf{u}$ associated with each initial cell type is 
$H^{(s)} =\sum_{j=1}^{\widetilde{N}(|\widetilde{\mathcal{C}}|-1)}\!\!\delta_{I(j),s} $, 
the number associated with each target cell type is 
$H^{(a)} = \sum_{j=1}^{\widetilde{N}(|\widetilde{\mathcal{C}}|-1)}\!\! \delta_{T(j),a}$, and
the number associated with each pair of cell types is 
$H^{(s,a)} = \sum_{j=1}^{\widetilde{N}(|\widetilde{\mathcal{C}}|-1)}\!\! \delta_{I(j),s} \delta_{T(j),a}$.

\begin{figure*}[h]
\centering
\includegraphics[width=\textwidth]{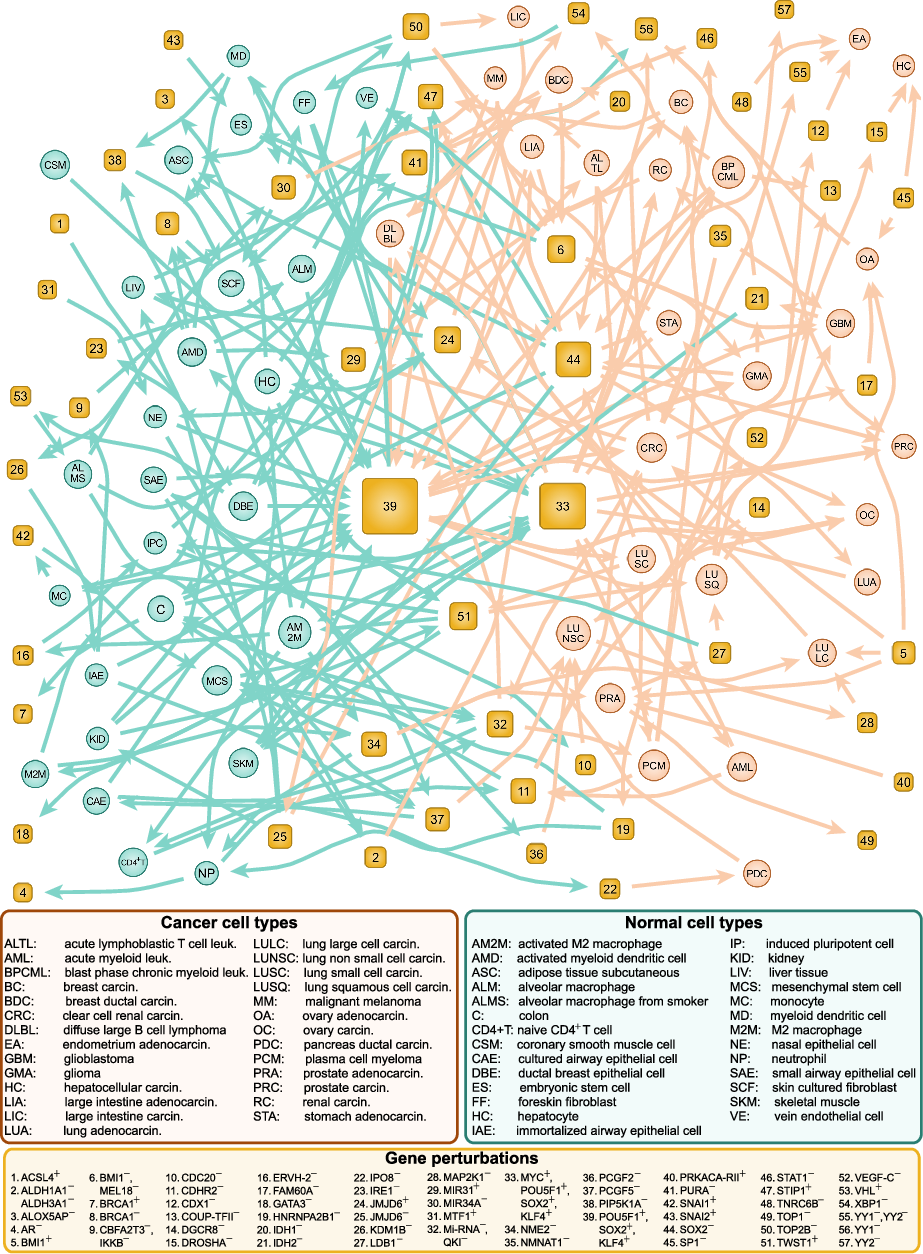}
\vspace{-1mm}
\caption[Bipartite network of gene perturbations overrepresented in  transdifferentiation transitions for the GeneExp dataset.]{
Counterpart of Fig.~\ref{fig:rnaseq_siggenes} for the GeneExp dataset. The circular node colors and shapes are as in Fig.~6.
}
\label{fig:geneexp_siggenes}
\end{figure*}

Then, the average inputs are:
\begin{equation}
 \bar{\mathbf{u}}^{(s,a)} = \frac{1}{H^{(s,a)}} \!\!\!\! \sum_{j=1}^{\widetilde{N}(|\widetilde{\mathcal{C}}|-1)} \!\!\!\!\mathbf{u}^{(j)} \delta_{I(j),s} \delta_{T(j),a}   \label{eq:pair_avg}
\end{equation}
for each pair of initial cell types $s$ and target cell types $a$.
The gene most strongly associated with each transition is  $v^{(s,a)} = \argmax_i \bar{u}_i^{(s,a)}$,
where $i \in \{1,...,|\mathcal{P}|\}$. 
 Defining $\mathbf{e}(i) = \big( e_{i'}(i) \big) = \big(\delta_{i,i'} \,|\, i' \in \{1,...,|\mathcal{P}|\}\big)$ to be
the unit (i.e., one-hot) vector associated with the $i$th perturbation,
we determine the genes most strongly associated with each initial cell type and  target cell type 
as  $\mathbf{v}^{(s)} = \sum_{a'\neq s} \mathbf{e}(v^{(s,a')})/(|\widetilde{\mathcal{C}}|-1)$ and 
$\mathbf{v}^{(a)} = \sum_{s'\neq a} \mathbf{e}(v^{(s',a)})/(|\widetilde{\mathcal{C}}|-1)$,  respectively.
From $\mathbf{v}^{(s)}$, we determine the probability that the observed number of occurrences $h^{(a)}$ of each
perturbation within $H^{(a)}$ states among a  target cell type $a$
exceeds the the multinomial distribution null hypothesis
\begin{equation}
P(X\geq h^{(a)}) = \sum_{n=h^{(a)}}^{H^{(a)}} \sum_{\{n_\ell\}}^{|\mathcal{P}|} \prod_{s\neq a}^{\widetilde{\mathcal{C}}} \begin{pmatrix} H^{(s,a)} \\ n_\ell \end{pmatrix} \left(v^{(s)}_\ell \right)^{n_\ell}, \label{supp_eq:exact_pvals}
\end{equation}
 where the middle sum is taken over all combinations $\{n_\ell\}$ such that $\sum_{\ell=1}^{|\mathcal{P}|} n_\ell = n$.
Exchanging $ s \leftrightarrow a $ in \cref{supp_eq:exact_pvals} yields the probability that the number of occurrences of a perturbation among an initial cell type is explained by the frequencies among the target cell types.
We apply the two-stage Benjamini-Hochberg multiple hypothesis correction to the $p$-values obtained from \cref{supp_eq:exact_pvals}  
at an false discovery rate of 1\% to obtain the significant genes associated with  transdifferentiation transitions
out of and into each cell type that are represented in \cref{fig:rnaseq_siggenes,fig:geneexp_siggenes}.

 \subsection*{Comparison with recent VAE methods}
Recent methods use VAEs to reconstruct the transcriptional states of perturbations applied to cell types when the post-perturbation transcriptional state is absent from the training data~\cite{Lotfollahi2019,Lotfollahi2023}.
We compare the performance of these methods in \cref{tab:VAE_comp} 
using a single-cell RNASeq dataset of interferon-$\beta$ stimulated and unstimulated peripheral blood mononuclear cells~\cite{Kang2018}. 
This dataset was previously used to demonstrate the efficacy of the VAE approach for the purpose of reconstructing transcriptional states in ref.~\citenum{Lotfollahi2019}.
We acquired the data and trained the VAE according to the documentation at \url{https://scgen.readthedocs.io/en/stable/installation.html}. 
We obtained the $R^2$ VAE estimates from the notebook file ``scgen\_perturbation\_prediction.ipynb'', available at \url{https://scgen.readthedocs.io/en/stable/tutorials/scgen\_perturbation\_prediction.html}.

The $R^2$ estimates for our method were computed using steps 1--4 in the subsection ``Defining perturbation responses'' above, with the following specifications: (i) $k=1$ since all data are from the same series of experiments and $p=$ interferon-$\beta$, (ii) all single-cell measurements of a given cell type were taken as replicates for the purposes of calculating $\langle \bar{\mathbf{X}}_{(m,p)}\rangle$ and $\langle \bar{\mathbf{X}}_{(m,0)}\rangle$, (iii) $\mathbf{B}_{(p)}$ is an average over the training cell types only. 
Using $m'$ to denote the test cell type, we then obtained $R^2$ between the actual state $\langle \bar{\mathbf{X}}_{(m',p)}\rangle$ and the predicted state $\langle \bar{\mathbf{X}}_{(m',0)}\rangle+\mathbf{B}_{(p)}$.
In this case,
the actual state is the transcriptional state of the stimulated test cell type and the predicted state is the sum
of the transcriptional state of the unstimulated test cell type and the average transcriptional response in the training cell types.

\subsection*{Data availability statement} 
Raw sequencing data is available through SRA~\cite{leinonen2010sequence} and GEO~\cite{Barrett2013}. Relevant accession numbers are included with the software and processed data for employing the method, which are available from GitHub at \url{https://github.com/twytock/cell\_reprogramming\_by\_transfer\_learning}.

\section*{Acknowledgements}
This work was supported by ARO grant No.\ W911NF-19-1-0383 and NIH/NCI grants No.\ P50-CA221747 (through the Malnati Brain Tumor Institute) and No.\ U54-CA193419 (through the Chicago Region Physical Sciences Oncology Center).
TPW also acknowledges support from NSF GRFP grant No.\ DGE-0824162 and NIH/NIGMS grant No.\ 5T32-GM008382.
The research benefitted from resources from NSF grant No.\ MCB-2206974
and the use of Quest High Performance Computing Facility at Northwestern University.
cd 
\section*{Author contributions}
TPW and AEM developed the research concept. TPW conducted the data collection, implemented the computer software, and analyzed the results. Both TPW and AEM wrote and edited the manuscript.

\bibliographystyle{pnas-new}
\bibliography{Model-Independent_Design}
\newpage
\pagebreak[4]
\clearpage

\beginsupplement
\makeatletter
\renewcommand*{\l@section}{\@dottedtocline{1}{0em}{2em}}
\renewcommand*{\l@subsection}{\@dottedtocline{2}{2em}{3em}}
\renewcommand*{\l@subsubsection}{\@dottedtocline{3}{5em}{3.5em}}
\makeatother
\begin{center}
\textbf{\LARGE Supplementary Material:\\ Cell reprogramming design by transfer learning of functional transcriptional networks}
\end{center}

\section*{Supplementary Figures}

\begin{figure}[h]
  \centering
  \includegraphics[width=85mm]{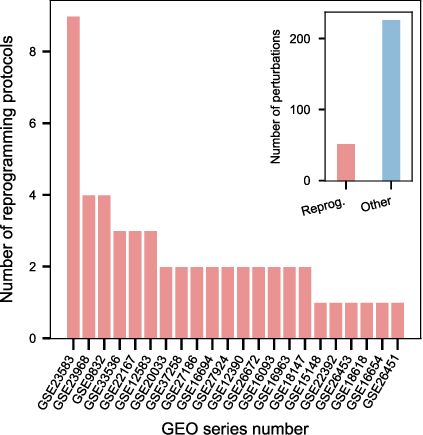}
 \caption{Number of unique reprogramming protocols in each GEO Series in the reprogramming partition of the GeneExp dataset.
 A protocol is unique if its initial cell type, final cell type, or set perturbed genes is different from all others in a given series. 
 \emph{Inset:} number of different perturbations in the dataset, which correspond to the number of protocols in the case of reprogramming.
  }
  \label{fig:num_strat}
\end{figure}

\newpage

\clearpage
\section*{Supplementary Tables}

\begin{table}[h]
\centering
\caption{\label{tab:VAE_comp}
 Comparison of VAE and linear models. {\rm  $R^2$ of the predicted versus actual gene expression response to interferon-$\beta$ stimulation in a single-cell RNA-seq dataset of peripheral blood monocyte cells~\cite{Kang2018}.  The predicted and actual transcriptional states are calculated using averages across all single-cell measurements for a given cell type.}}
\begin{tabular}{l r r r}
\hline
Test cell type & $R^2_{\rm VAE}$ & $R^2_{\rm Linear}$ & $R^2_{\rm VAE}-R^2_{\rm Linear}$  \\[2pt] \hline
CD4T &                 0.963 & 0.904 &      0.059 \\
FCGR3A+Mono & 0.939 & 0.916 &      0.023 \\
CD14+Mono &      0.925 & 0.873 &      0.052 \\
B &                        0.855 & 0.935 & $-$0.080 \\
NK &                     0.867 & 0.918 & $-$0.051 \\
CD8T &                 0.930 & 0.907 &      0.023 \\
Dendritic &            0.958 & 0.939 &      0.019 \\ \hline
Average &             0.920 & 0.913 &      0.007 \\ \hline
\end{tabular}
\end{table}

\begin{table}[h]
\centering
\caption{\label{tab:RNAseq_pert}
{\rm Metadata on cell type, gene perturbation, and treatment for the RNA-Seq data not included with GTEx. Available in the ``data'' folder of the GitHub repository at} \protect\url{https://github.com/twytock/cell_reprogramming_by_transfer_learning}.}

\end{table}

\begin{table}[h]
\centering
\caption{\label{tab:gexp_exps} {\rm Metadata on cell type, gene perturbation, and treatment for the GeneExp dataset. Available in the ``data'' folder of the GitHub repository at }\protect\url{https://github.com/twytock/cell_reprogramming_by_transfer_learning}.}

\end{table}

\end{document}